\documentclass[%
 reprint,
nofootinbib,
 amsmath,amssymb,
 aps,
prstab,
]{revtex4-2}

\usepackage{graphicx}
\usepackage{dcolumn}
\usepackage{bm}

\usepackage{multirow}
\usepackage{gensymb}

\begin{document}

\preprint{APS/123-QED}

\title{HL-LHC layout for fixed-target experiments in ALICE based on crystal-assisted beam halo splitting}

\author{Marcin Patecki}
\email{Marcin.Patecki@pw.edu.pl}

\affiliation{Warsaw University of Technology, Faculty of Physics, ul. Koszykowa 75, 00-662 Warsaw, Poland. }
\author{Alex Fomin}%
\author{Daniele Mirarchi}
\author{Stefano Redaelli}
\affiliation{%
 European Organization for Nuclear Research (CERN), CH-1211 Geneva 23, Switzerland
}
\author{Cynthia Hadjidakis}
\affiliation{Universit\'e Paris-Saclay, CNRS/IN2P3, IJCLab, 91405 Orsay, France}

\author{Francesca Galluccio}
\affiliation{INFN Sezione di Napoli, Complesso Universitario di Monte Sant’Angelo, Via Cintia, 80126 Napoli, Italy }

\author{Walter Scandale}
\affiliation{Blackett Laboratory, Imperial College, London SW7 2AZ, UK}
\affiliation{INFN Sezione di Roma, Piazzale Aldo Moro 2, 00185 Rome, Italy}

\date{\today}

\begin{abstract}
The Large Hadron Collider (LHC) at the European Organization for Nuclear Research (CERN) is the world's largest and most powerful particle accelerator colliding beams of protons and lead ions at energies up to 7~Z~TeV, Z is the atomic number. ALICE is one of the detector experiments optimised for heavy-ion collisions.
A fixed-target experiment in ALICE is being considered to collide a portion of the beam halo, split using a bent crystal inserted in the transverse hierarchy of the LHC collimation system, with an internal target placed a few meters upstream of the existing detector. This study is carried out as a part of the Physics Beyond Collider effort at CERN.
Fixed-target collisions offer many physics opportunities related to hadronic matter and the quark-gluon plasma to extend the research potential of the CERN accelerator complex. 
Production of physics events depends on the particle flux on target.
The machine layout for the fixed-target experiment is developed to provide a flux of particles on the target high enough to exploit the full capabilities of the ALICE detector acquisition system.
This paper summarises the fixed-target layout consisting of the crystal assembly, the target and downstream absorbers. We discuss the conceptual integration of these elements within the LHC ring, the impact on ring losses,
 and expected performance in terms of particle flux on target.
\end{abstract}

\maketitle

\section{Introduction}
\label{s:intro}
Advancements in the knowledge of fundamental constituents of matter and their interactions are usually driven by the development of experimental techniques and facilities, with a significant role of particle accelerators.
The Large Hadron Collider (LHC)~\cite{LHC} at the European Organization for Nuclear Research (CERN) is the world's largest and most powerful particle accelerator colliding opposite beams of protons (p) and lead ions (Pb), allowing for unprecedentedly high centre-of-mass energies of up to 14~TeV and 5.5~TeV per nucleon, respectively.
An ALICE fixed-target (ALICE-FT) programme~\cite{A-FT_proposal} is 
being considered to extend the research potential of the LHC and the ALICE experiment~\cite{ALICE}. The setup of \mbox{in-beam} targets at the LHC is particularly challenging because of the high-intensity frontier of LHC beams~\cite{PBC-report}.

Fixed-target collisions in the LHC are designed to be operated simultaneously with regular head-on
collisions without jeopardising the LHC efficiency during its main p-p physics programme. Several unique advantages are offered with the fixed-target
mode compared to the collider mode. With a high density of targets, high yearly luminosities can be easily
achieved, comparable with luminosities delivered by the LHC (in the collider mode) and Tevatron~\cite{AFTER1}. In
terms of collision energy, the \mbox{ALICE-FT} layout would provide the most energetic beam ever in the
fixed-target mode with the centre of mass energy per nucleon-nucleon of 115 GeV for proton beams and
72 GeV for lead ion beams~\cite{AFTER1}, in between the nominal Relativistic Heavy Ion Collider (RHIC) and Super
Proton Synchrotron (SPS) energies. Thanks to the boost between the colliding-nucleon centre-of-mass
system and the laboratory system, access to far backward regions of rapidity is possible with the ALICE
detector, allowing to measure any probe even at far ranges of the backward phase space, being utterly
uncharted with head-on collisions~\cite{AFTER1}. Moreover, the possibility of using various species of the target
material extends the variety of physics cases, especially allowing for unique neutron studies~\cite{AFTER1}.
The physics potential~\cite{AFTER1, A-FT_proposal} of such a \mbox{fixed-target} programme covers an intensive study of strong interaction
processes, quark and gluon distributions at high momentum fraction (x), sea quark and heavy-quark content
in the nucleon and nucleus and the implication for cosmic ray physics. The hot medium created in
ultra-relativistic heavy-ion collisions offers novel quarkonium and heavy-quark observables in the energy
range between the Super Proton Synchrotron (SPS) and the Relativistic Heavy Ion Collider (RHIC), where
the QCD phase transition is postulated.

A significant innovation of our proposal is to bring particles of high energy collider to collisions with a \mbox{fixed-target} by splitting a part of the beam halo using a bent crystal. Particles entering the crystal with a small impact angle 
($\rm \leq 2.4~\mu rad$ 
for silicon crystals and proton energy of 7~TeV~\cite{Daniele_thesis}) 
undergo the channeling process resulting in a trajectory deflection equivalent to the geometric bending angle of the crystal body~\cite{cry_book, cry_channeling1}.
Such a setup enables an in-beam target at a safe distance from the circulating beam. 
This type of advanced beam manipulation with bent crystals builds on the experience accumulated in different accelerators (see for example
~\cite{scandale2019, scandale2022}), and in particular, on the successful results achieved in the \mbox{multi-TeV} regime for beam collimation at the LHC~\cite{cry_coll1, cry_coll3, cry_coll4}. The halo-splitting technique allows profiting from the circulating
beam halo particles that are not contributing to the luminosity production and are typically disposed of by the collimation system. 

The problem that we address is to design the machine layout that
provides a number of protons on the target high enough to exploit the full capabilities of the ALICE detector
acquisition system without affecting the LHC availability for regular beam-beam collisions. Our proposal of the ALICE-FT layout~\cite{HB2021} follows general guidelines on technical feasibility and impact on the LHC accelerator of potential fixed-target experiments provided
by the LHC Fixed Target Working Group of the CERN Physics Beyond Colliders forum~\cite{PBC, PBC-report}. We also profit from the  preliminary designs reported in~\cite{PBC_fran, PBC_alex} and  from the design study of an analogous fixed target experiment at the LHC proposed to measure electric and magnetic dipole moments of short-lived baryons~\cite{Daniele_LHCb}.  

In this paper, we summarise the ALICE-FT machine layout. We report on the conceptual integration of its elements (crystal and target assemblies, downstream absorbers), their impact on ring losses, and expected performance in terms of particle flux on target. We also discuss a method of increasing the flux of particles on the target by setting the crystal at the optimal betatron phase by applying 
a local optics modification in the insertion hosting the ALICE experiment (IR2). This method is independent of the crystal location, allowing for a crystal installation in a place with good space availability. Moreover, it allows to recover the maximum performance of the system in case of changes in beam optics in the LHC.

\section{Machine configuration}
A potential installation of the ALICE-FT setup 
will coincide with a major LHC upgrade in terms of instantaneous luminosity, commonly referred to as the High-Luminosity LHC (HL-LHC)~\cite{HL-LHC}, taking place in the Long Shutdown~3 \mbox{(2026-2028)}, to make it ready for the LHC Run~4 starting in 2029. Some of the expected beam parameters, having a direct impact on the ALICE-FT experiment performance, are given in Table~\ref{t:beam_params}.
One key beam parameter to be upgraded is the total beam current that will increase nearly by a factor of two, up to about 1.1~A, leading to  about 0.7~GJ of total beam energy stored in the machine.
\begin{table}[htb]
\caption{Some proton-beam parameters of the future HL-LHC beams important for the ALICE-FT experiment, referred to as \textit{standard} in \cite{HL-LHC}.}
\label{t:beam_params}
\centering
   \def\arraystretch{1.2}
\begin{tabular}{l  l  l}
\hline
\hline
Colliding-beam energy    &$\rm E$                 & 7~TeV    \\
Bunch population            &$\rm N_{b}$ & $\rm2.2 \cdot 10^{11}$   \\
Maximum number of bunches           &$\rm n_b$                     & 2760  \\
Beam current                &I                      & 1.09~A  \\
Transverse normalised emittance & $\rm \varepsilon_{n}$&  2.5~$\rm [\mu m]$\\
$\rm \beta^{*}$ at IP2      &                           & 10  m  \\ 
Beam crossing angle at IP2  &                & 200~$\rm \mu$rad   \\
\hline
\hline
\end{tabular}
\end{table}
A~highly efficient collimation system is therefore required in the LHC~\cite{coll_system} to protect its elements, especially the superconducting magnets, from impacts of particles from the beam. 
The collimation system is organised 
in a precise multi-stage transverse hierarchy (see Table~\ref{t:coll_system}) over two dedicated insertions (IRs): IR3 for momentum cleaning and IR7 for betatron cleaning. Each collimation insertion features a three-stage cleaning based on primary collimators (TCPs), secondary collimators (TCSGs) and absorbers (TCLAs). In addition,  dedicated collimators are present in specific locations of the ring to provide protection of sensitive equipment (e.g. TCTP for the inner triplets), absorption of physics debris (TCL) and beam injection/dump protection (TDI/TCDQ-TCSP). The collimation system undergoes an upgrade, as described in~\cite{HL-LHC_coll}, to make it compatible with HL-LHC requirements, but the general working principle will remain the same. The system is designed to sustain beam losses up to 1~MJ without damage and with no quench of superconducting magnets.
%
\begin{table}[htb]
\caption{HL-LHC collimation settings expressed in units of RMS beam size ($\rm \sigma$), assuming a Gaussian beam distribution and transverse normalised emittance $\rm \varepsilon_{n} = 2.5~\mu m$.} 
\label{t:coll_system}
\centering
   \def\arraystretch{1.2}
\begin{tabular}{l l l }
\hline
\hline
Coll. family        & IR        & Settings ($\rm \sigma$)   \\
\hline
TCP/TSCG/TCLA       & 7         & 6.7/9.1/12.7              \\
TCP/TSCG/TCLA       & 3         & 17.7/21.3/23.7            \\
TCTP                & 1/2/5/8   & 10.4/43.8/10.4/17.7       \\
TCL                 & 1/5       & 14.2                      \\
TCSP/TCDQ           & 6         & 10.1/10.1                 \\
\hline
\hline
\end{tabular}
\end{table}

The halo-splitting scheme relies on placing a crystal into the 
transverse hierarchy of the betatron 
collimation system, in between the primary and secondary stage of IR7 collimators, such that the collimation system efficiency is not affected. Placing the splitting crystal closer to the beam than the primary collimators would not be possible without designing a downstream system capable of withstanding the collimation design loss scenarios. Retracting the crystal at larger amplitudes avoids this problem while still allowing intercepting a significant fraction of the multi-stage halo, as shown below. In this scheme, 
a fraction of secondary halo particles redirected toward the target can be used for fixed-target collisions instead of disposing them at the absorbers, in a safe manner. Note that, for this scheme to work, the crystal does not need to be installed in IR7. In fact, the halo splitting is done in the vicinity of the experiment, where the target needs to be located.

\section{ALICE-FT layout}

\begin{figure}[htb]
    \centering
    \includegraphics[width=0.49\textwidth]{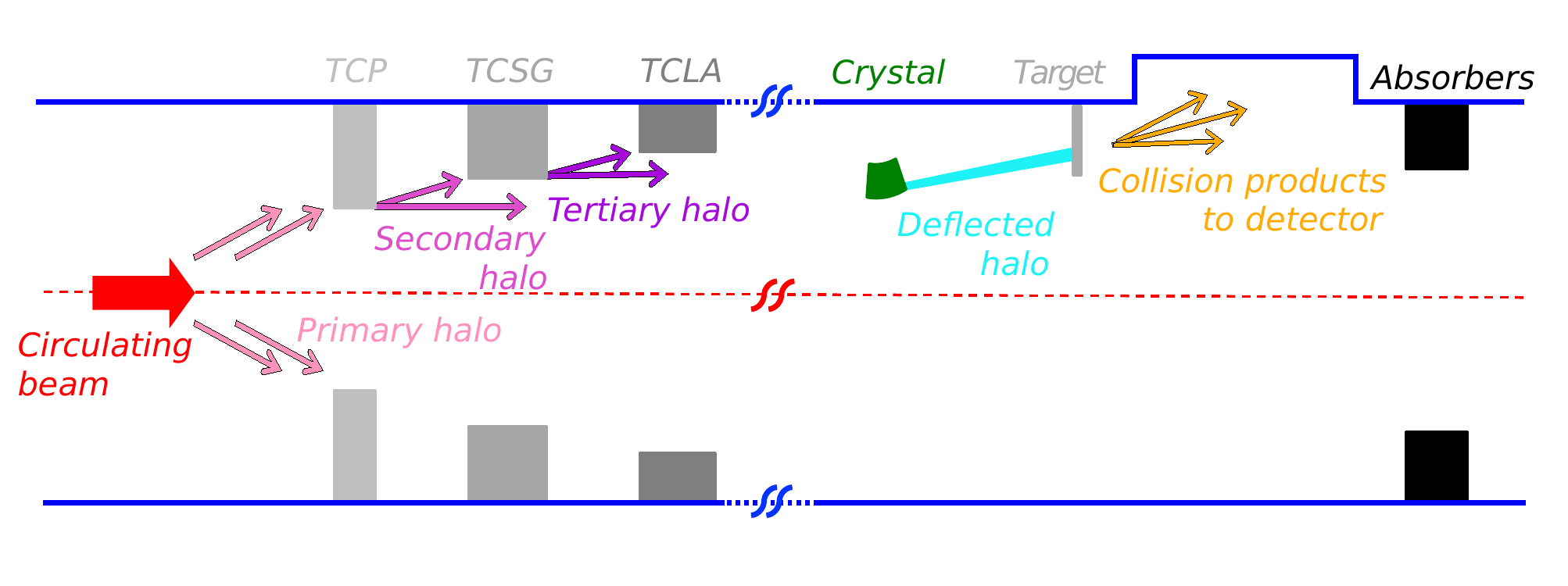}
    \caption{Working principle of the crystal-based fixed-target experiment (right side of the graphics) being embedded into the multi-stage collimation system (left side of the graphics). Graphics based on~\cite{Daniele_LHCb}, mostly by D.~Mirarchi.
    }
    \label{fig:FT_layout}
\end{figure}

A general concept of the ALICE-FT layout is illustrated in Fig.~\ref{fig:FT_layout}.
A fraction of secondary halo particles are intercepted by the crystal and steered toward the target. Collision products are registered by the ALICE detector, which can handle in the order of $\rm 10^{7}$ protons on target per second~\cite{AFTER1}. Possible losses originating either from the crystal or from the target are intercepted by absorbers which would be installed downstream of the detector.

The ALICE-FT will act on B1 due to ALICE detector geometry. We assume the layout to be operated with the optics envisioned for the \mbox{HL-LHC} (version 1.5~\cite{optics_v1.5} at the moment), although it is not excluded that more optimized optics could be envisaged if needed. Since ALICE operates at low luminosities in proton runs, typically with offset levelling, this insertion does not demand tight optics constraints like the high-luminosity insertions for the general-purpose detectors \cite{HL-LHC2}.
Actually, a minor, local modification of the optics in the IR2 region is proposed to enhance the system performance by optimising the betatron oscillation phase at the location of the crystal. This will not affect the rest of the machine. 

Safe integration of the crystal into the hierarchy of the collimation system is achieved by setting the crystal in the shadow of IR7 primary collimators. As discussed in~\cite{Daniele_LHCb}, the relative retraction of the crystal with respect to the IR7 primary collimator should not be smaller than $\rm 0.5~\sigma$ (RMS beam size), mostly to account for optics and orbit errors that, if not under control, could cause the crystal to become primary collimator accidentally. The retraction should be kept as low as possible to maximise the number of protons impacting the target~\cite{PBC-report} as shown in Fig.~\ref{fig:cry_nsig}. Similarly to~\cite{Daniele_LHCb}, we also assume that, for machine safety reasons, the distance from the deflected beam to the aperture and the distance from the target to main beams should be at least 4~mm. The system is to be installed in the vertical plane in order to avoid issues related to the beam dump system operating in the horizontal plane: this is the plane of the dump kickers and is subject to fast losses in case of over-populate abort gap or in case of (unlikely) dump failures~\cite{dump1, dump2}. In the case of a horizontal setup, larger aperture margins on settings would be applied to lead to lower rates of impinging halo, which we prefer to avoid. 

\begin{figure}[!h]
    \centering
    \includegraphics[width=0.49\textwidth]{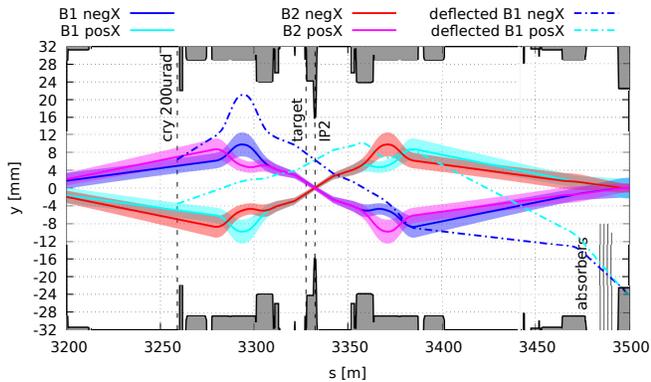}
    \caption{The proposed layout of the ALICE-FT experiment. Both beams (B1 and B2) with their envelopes ($\rm 7.3~\sigma$) are given with solid lines for both ALICE solenoid polarities (posX and negX). Deflected beams are given in dashed blue lines. The machine aperture is given in solid black lines. Vertical dashed lines mark the locations of crystals, target and IP2, respectively. The location of absorbers is marked in the right bottom corner.}
    \label{fig:FT_layout_IR2}
\end{figure}

\begin{figure}[]
    \centering
    \includegraphics[width=0.49\textwidth]{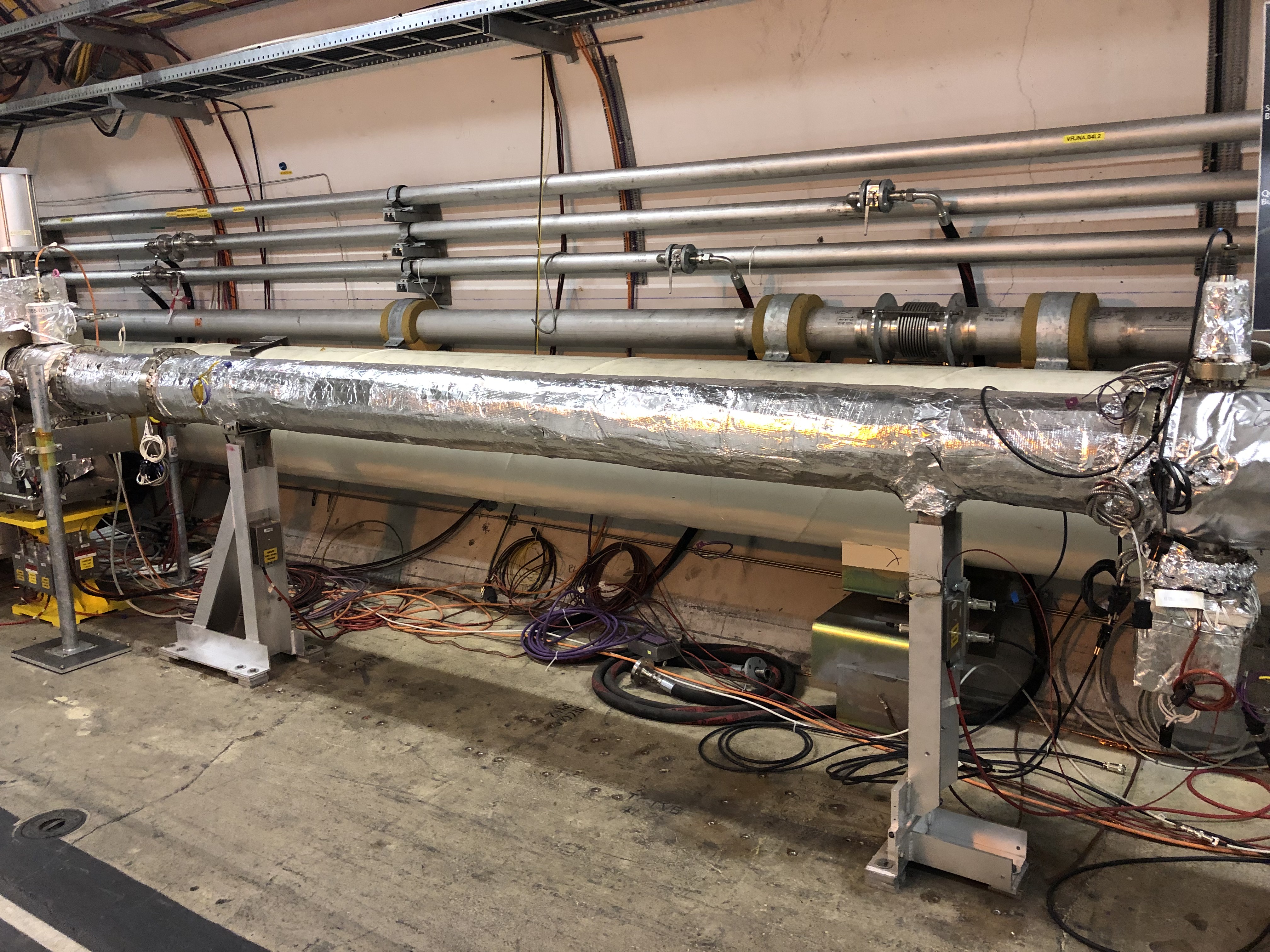}
    \caption{Conditions inside the LHC tunnel at the proposed location for an installation of the crystal.}
    \label{fig:tunnel_photo}
\end{figure}

The beams crossing scheme at IP2 is also in the vertical plane. Furthermore, the main solenoid of the ALICE detector can be operated in two polarities which affects the slope of both beams at IP2. We mark the negative slope of the LHC beam 1 (B1) at IP2 as \textit{negX} and the positive slope as \textit{posX}.
Given the trajectory of the B1 for both crossing schemes (\textit{negX} and \textit{posX}), aperture restrictions and space availability, the optimal longitudinal coordinate for the crystal installation is 3259m (with 0 at IP1) as it allows for having just one crystal serving both crossing schemes. 
The graphical illustration of the proposed layout, which fulfils all the above requirements, is given in Fig.~\ref{fig:FT_layout_IR2} and conditions inside the LHC tunnel are depicted in Fig.~\ref{fig:tunnel_photo}. 

We consider a 16-mm long crystal made of silicon, with 110 bending planes and a bending radius of 80~m, the same as a bending radius of crystals already used in the LHC, following the parametric studies reported in~\cite{cry_coll3} to ensure an optimum crystal channelling performance at the LHC top energy while keeping the nuclear interaction rate as low as possible.
The crystal bending angle of $\rm 200~\mu rad$ is chosen to serve both crossing schemes and to fit the deflected beam within the available aperture. Any upstream location of the crystal would require having two crystals with different bending angles, depending on the crossing scheme.

The target assembly is planned to be installed nearly 5~m upstream from IP2 with a target made of either light or heavy material,~e.g. carbon or tungsten of about 5~mm in length.
Details on target design studies can be found in~\cite{target_design}.

Four absorbers, of the same design as collimators already used in the LHC, are proposed to be installed about 150~m downstream from the IP2. The first three are proposed to be made of 1~m long carbon-fibre-carbon composite jaws, as the present \mbox{TCSGs} in the LHC, while the last one is to be made of 1~m long tungsten jaws, as the present TCLAs in the LHC; similarly as in~\cite{Daniele_LHCb}. The difference is that in our study, we use a large opening of about 50~$\rm \sigma$ that still intercepts the channelled beam while being well in the shadow of the entire collimation system. Such a choice is driven by minimising the impact of these extra absorbers on the regular collimation system and machine impedance instead of searching for minimum gaps that maintain the collimation hierarchy. As will be shown later, we do not experience any cleaning-related issues. The proposed setup of absorbers follows a performance-oriented approach with the potential to reduce the number of required absorbers, based on the energy deposition study to be done in the future. 

\section{Expected performance}
The MAD-X code~\cite{madx} is used to manage the HL-LHC model, to prepare suitable lattice and optics descriptions used as input to tracking studies, and to calculate the trajectory of particles experiencing an angular kick equivalent to the crystal bending angle. Detailed evaluation of the layout performance is done using multi-turn particle tracking simulations in SixTrack~\cite{sixtrack1} that allows a simplectic, fully chromatic and 6D tracking along the magnetic lattice of the LHC, including interactions with collimators and bent crystals, and a detailed aperture model of the machine~\cite{sixtrack2, sixtrack3}. In our simulations, we use at least two million protons, initially distributed over a narrow ring of radius $\rm r+dr$ slightly above 6.7~$\rm \sigma$ in the normalised transverse vertical \mbox{position-angle} phase space (y,y') which allows an estimation of the number of protons impacting the collimation system (including the crystal and the target of the ALICE-FT layout) as well as the density of protons lost per metre in the aperture with a resolution of 10~cm along the entire ring circumference.

\begin{figure}[!h]
    \centering
    \includegraphics[width=0.49\textwidth]{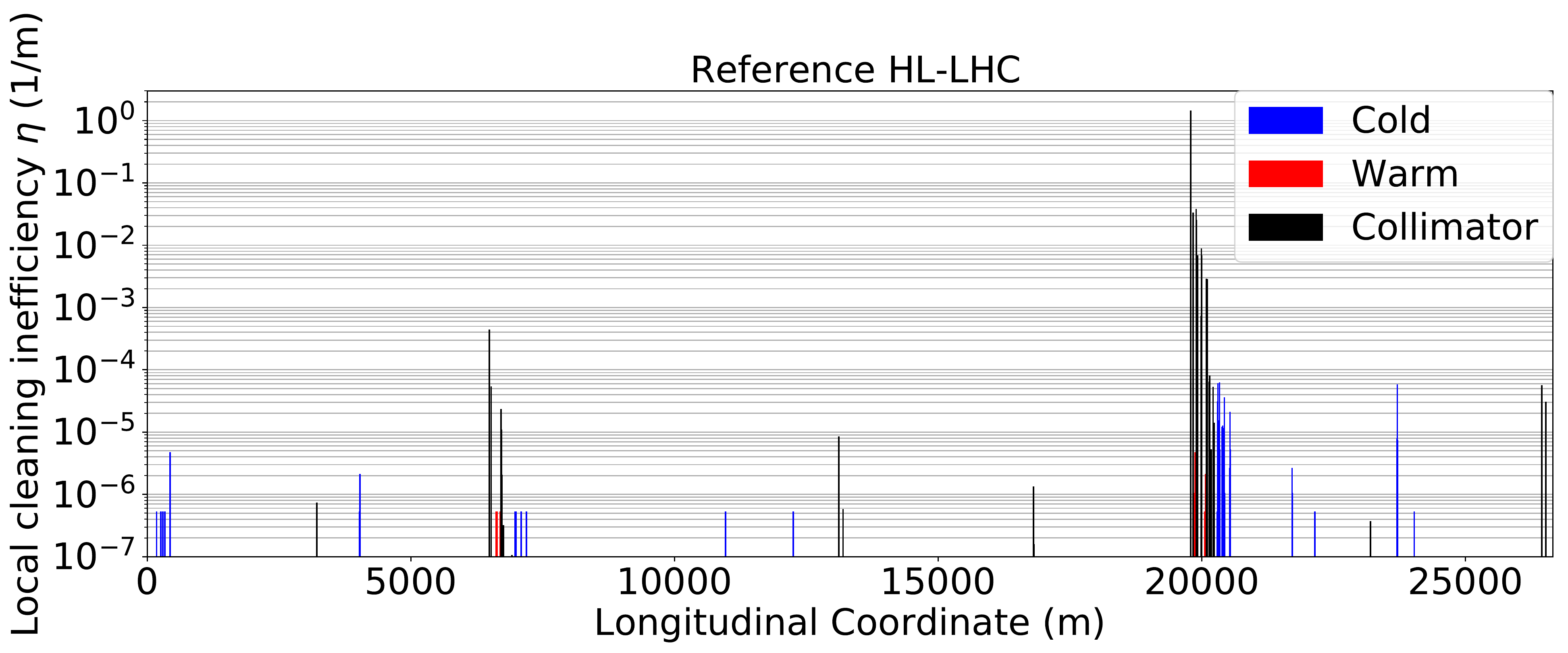}
    \includegraphics[width=0.49\textwidth]{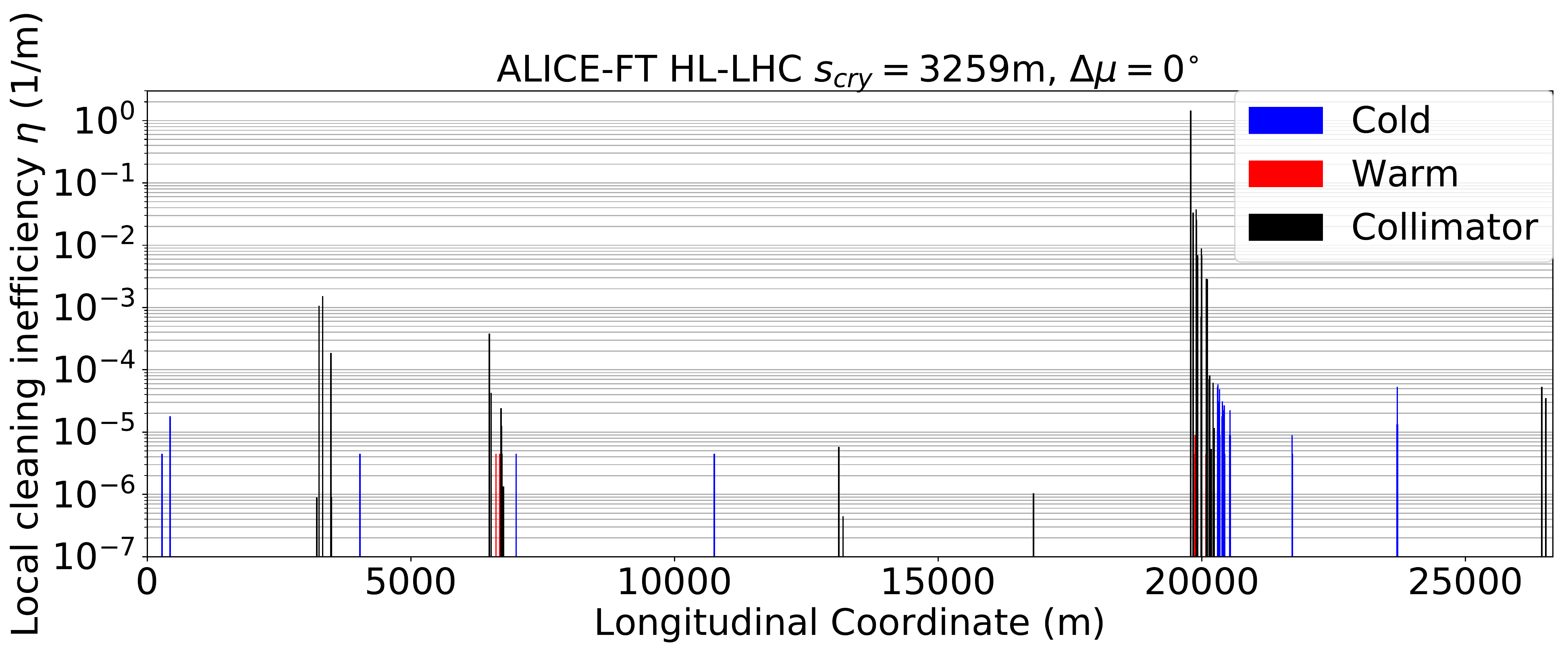}
    \includegraphics[width=0.49\textwidth]{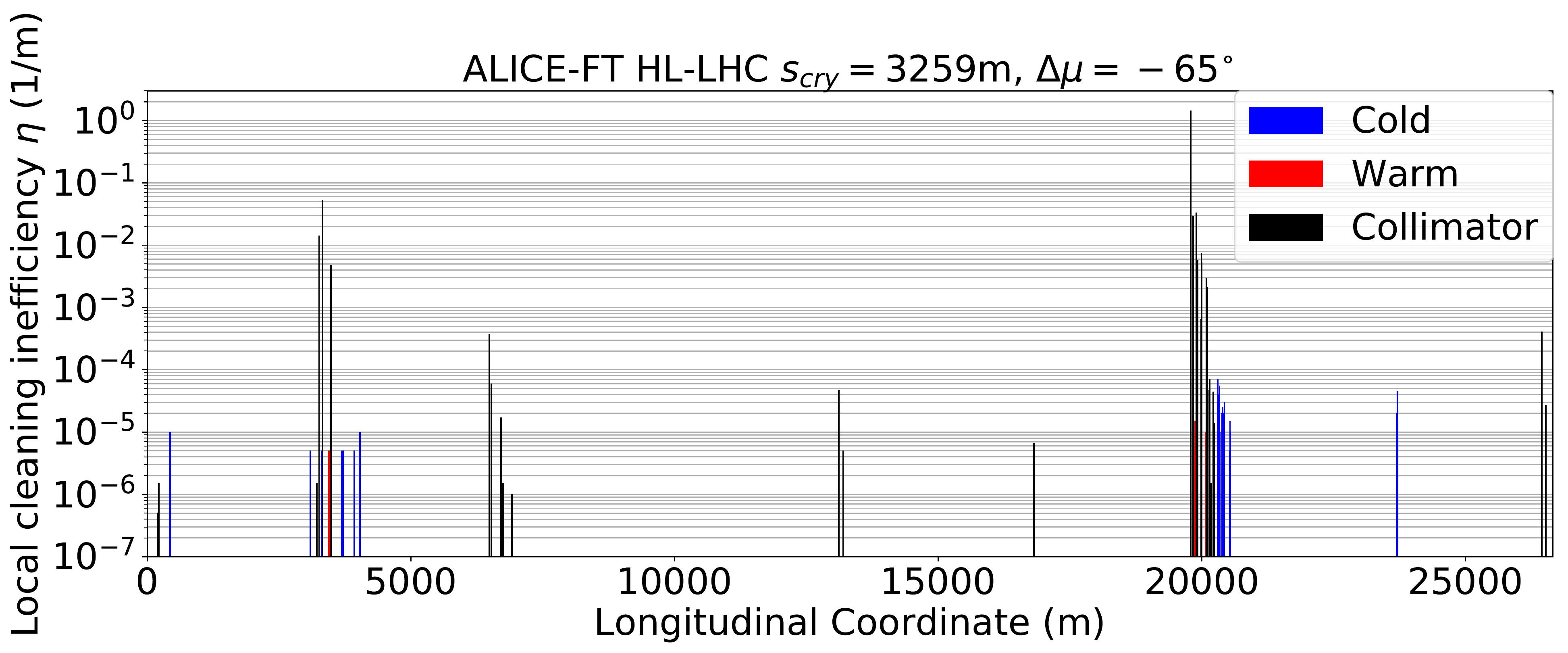}
    \caption{Comparison of loss maps for the machine without (top) the ALICE-FT system, with the ALICE-FT system without the optics optimisation (middle) and with the ALICE-FT system with the optics optimisation (bottom). The local cleaning inefficency (vertical axis) is a measure of the number of protons not intercepted by the collimation system and impacting the machine aperture. The simulation limit of 1 proton lost in the machine aperture corresponds to $\rm 5 \cdot 10^{-7} m^{-1}$ in a 10~cm longitudinal bin. No abnormal loss spikes are present in the loss maps.}
    \label{fig:loss_maps}
\end{figure}

\begin{figure}[!h]
    \centering
    \includegraphics[width=0.49\textwidth]{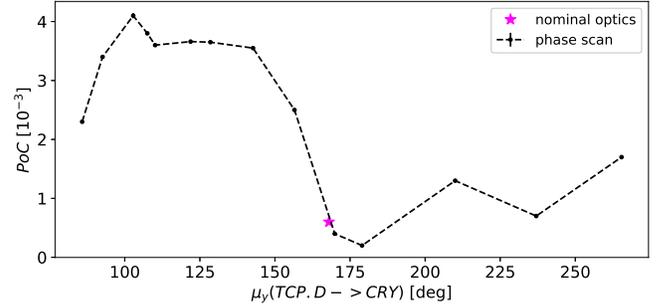}
    \caption{The dependence of the fraction of the number of protons impacting the crystal over the number of protons impacting the primary collimator (PoC)
    on the betatron phase from the primary vertical collimator in IR7 (TCP.D) to the IR2 crystal (CRY). Crystal retraction is $\rm 7.9~\sigma$. Statistical errors are in the order of 1\%, which makes the bars hardly visible.}
    \label{fig:phase_scan}
\end{figure}

\begin{figure}[!h]
    \centering
    \includegraphics[width=0.49\textwidth]{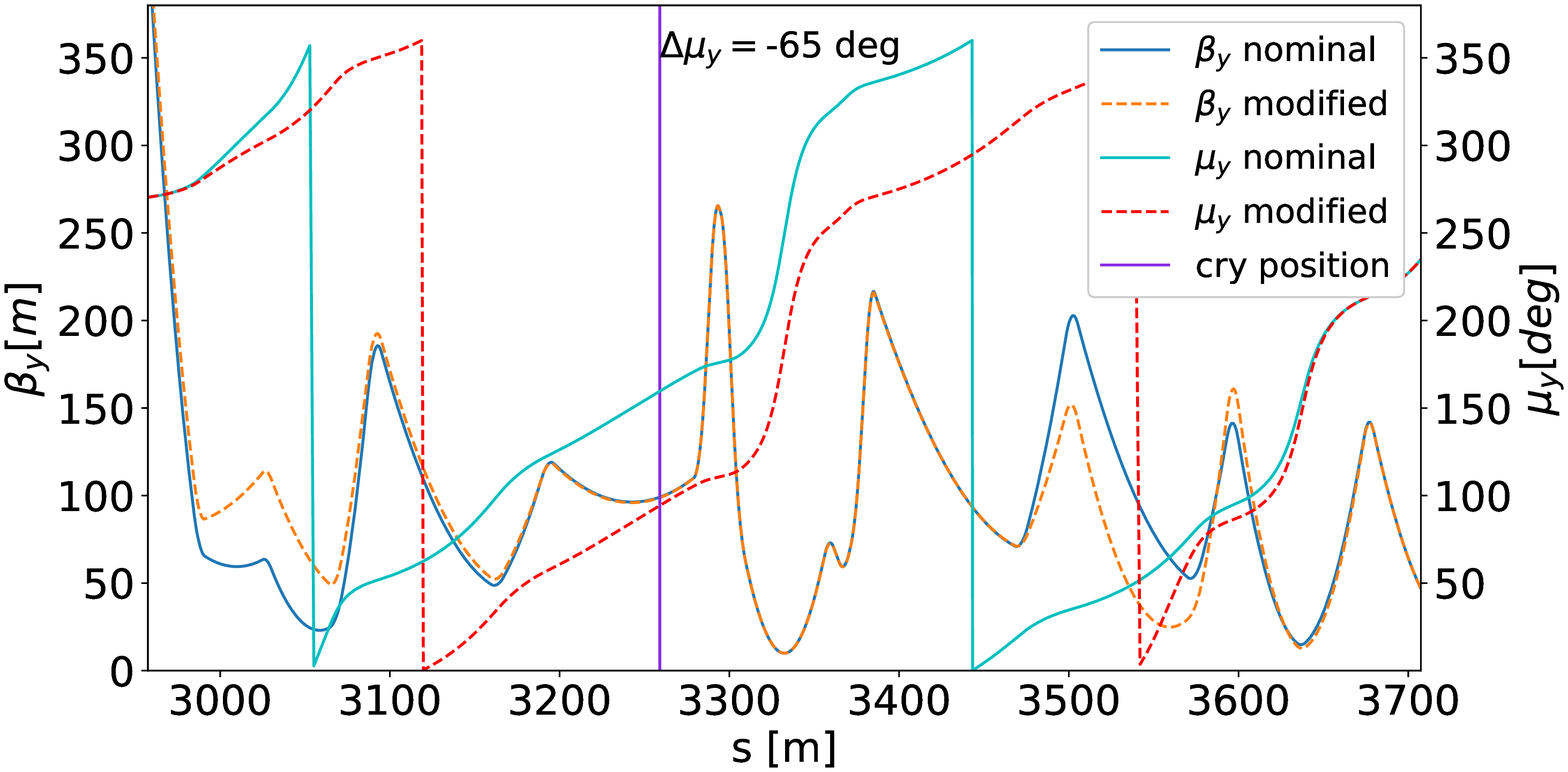}
    \caption{Vertical $\rm \beta$ function and betatron phase for nominal (solid lines) and modified (dashed lines) optics. Position of the crystal is marked with a vertical purple line.}
    \label{fig:optics_modified}
\end{figure}

The ALICE-FT experiment must be compatible with the standard physics programme of the LHC, meaning that it cannot add any operational limitations, mostly related to particle losses, which must stay within acceptable limits.
This is demonstrated in Fig.~\ref{fig:loss_maps} where a loss map of the machine including the ALICE-FT system does not contain any abnormal loss spikes comparing to the reference loss map of the machine without the \mbox{ALICE-FT} system. The only new spikes correspond to protons impacting the elements of the ALICE-FT setup.
The setup is developed to provide as many protons as possible impacting the crystal~($\rm N_{PoC}$) in order to maximise, for a given channeling efficiency, the number of protons on target. For a given crystal retraction relative to primary collimators, the proton flux on the crystal strongly depends on the betatron oscillation phase advance between the primary collimators at IR7 and the IR2 crystal. With the nominal optics, the phase advance is nearly least favourable, leading to a low $\rm N_{PoC}$. This can be corrected by a minor, local modification of the IR2 optics, implemented by changing strengths of IR2 quadrupole magnets labelled with natural numbers from 4 to 10 (lower the number, closer the magnet to the IP2) on both sides of the IP2. Quadrupoles upstream of the IP2 were constrained to shift the phase advance at the crystal while keeping the IP2 optics parameters unchanged. Quadrupoles downstream of the IP2 were constrained to recover the same optical parameters, including the betatron phase, as in the nominal optics.
The maximum $\rm N_{PoC}$ was found for phase advance shifted by $\rm -65^{\circ}$, as shown in Fig.~\ref{fig:phase_scan}. 
The corresponding optical $\rm \beta_{y}$ function in IR2 is given in Fig.~\ref{fig:optics_modified} and changes in strengths of quadrupoles are summarised in Table~\ref{t:quads_strength}. Such a modification of optics is well feasible to be implemented at the IR2 and does not affect the rest of the machine, especially losses along the ring as shown in Fig.~\ref{fig:loss_maps}.  
\begin{table}[h]
\caption{Normalised strengths of quadrupoles for nominal and modified optics. IR2 left and IR2 right stand for regions upstream and downstream from the IP2, respectively.} 
\label{t:quads_strength}
\centering
   \def\arraystretch{1.2}
\begin{tabular}{l c c c c}
\hline
\hline
    & \multicolumn{4}{c}{Quadrupole strength $\rm k_{1}$ [$\rm 10^{-3}~m^{-2}$]} \\
Quadrupole    & \multicolumn{2}{c}{IR2 left} & \multicolumn{2}{c}{IR2 right} \\ 
number         & nominal   & modified     & nominal   & modified \\
10                  & -6.39     & -6.15     &  7.30     &  7.30 \\
9                   &  7.01     &  6.89     & -6.60     & -6.82 \\
8                   & -5.41     & -3.59     &  6.71     &  6.30 \\
7                   &  7.60     &  7.42     & -6.36     & -7.47 \\
6                   & -4.91     & -4.17     &  4.33     &  4.20 \\
5                   &  2.99     &  2.88     & -3.63     & -4.09 \\
4                   & -2.80     & -2.67     &  3.74     &  2.60 \\
\hline
\hline
\end{tabular}
\end{table}

\begin{figure}[!hbt]
    \centering
    \includegraphics[width=0.49\textwidth]{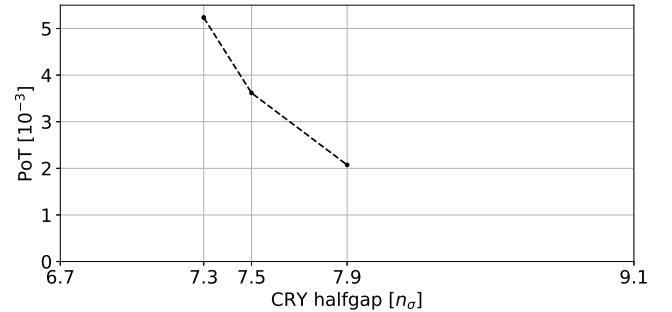}
    \caption{Fraction of the number of protons impacting the target over the number of protons impacting the primary collimator (PoT) for some values of crystal half-gap expressed as the number of beam $\rm sigma$. Limits of the horizontal axis correspond to half-gaps of primary and secondary collimators in IR7. Statistical errors are in the order of 1\%, which makes the bars hardly visible.}
    \label{fig:cry_nsig}
\end{figure}

The number of protons hitting the target depends on the number of protons hitting the crystal, their \mbox{phase-space} distribution at the crystal entry face, parameters of the crystal and crystal position and orientation inside the beam pipe. All these phenomena are treated by the simulation code we use. Protons hitting the crystal emerge from the collimation system. Therefore, their number and phase-space distribution are subject to a complex multi-turn process which is also simulated. As a result, we obtain PoC and PoT being fractions of protons hitting the crystal or the target (respectively) over all protons intercepted by the collimation system. Three values of PoT resulting from performed simulations and depending on the relative crystal retraction are given in Fig.~\ref{fig:cry_nsig}.

 These scaling factors allow us to estimate the actual number of protons hitting the crystal and (more interestingly) the target under realistic conditions of the beam, where the most important factors are beam intensity and lifetime. 
 An exponential decrease of the beam intensity is assumed, characterised by the time coefficient $\rm \tau$ interpreted as a beam lifetime, which depends on beam parameters and machine state. A dominant contribution to the total beam lifetime comes from the beam burn-off due to collisions~($\rm \tau_{BO}$), while the number of protons on the target $\rm N_{PoT}$ depends mostly on $\rm \tau_{coll}$ corresponding to beam core depopulation towards tails  that are intercepted by the collimation system.
Following the same assumptions as in~\cite{Daniele_LHCb}, 
with $\rm I_{0}$ being the initial beam intensity, time coefficients $\rm \tau_{BO} \approx 20h$ 
and $\rm \tau_{coll} \approx 200h$,  
the number of protons impacting the target per 10~h long fill ($\rm T_{fill}$) in 2018 operation conditions can be estimated as:

\begin{equation}
\label{eq:NPoT}
\begin{split}
\rm N_{PoT} &= \frac{1}{2}PoT \int_{0}^{T_{fill}} \frac{I_{0}}{\tau_{coll}}exp\left(-\frac{t}{\tau_{BO}}\right)exp\left(-\frac{t}{\tau_{coll}} \right)dt \\
&\approx 2.7 \times 10^{10}.
\end{split}
\end{equation}
This corresponds to an average flux of protons on target of $\rm 7.5 \times 10^{5}$~p/s being roughly one order of magnitude away from the design goal of $\rm 10^{7}$~p/s. These two numbers are summarised in Table~\ref{t:flux}. Please also note that the intensity of HL-LHC beams will be about a factor of two larger, which most probably will result in a larger proton flux on target, much closer to the design goal. However, the final assessment can be established more precisely by studying the lifetime and beam loss performance of HL-LHC beams.
\begin{table}[htb]
\caption{Expected proton flux on target based on 2018 operation conditions compared with expected capabilities of ALICE detector acquisition system. The first number may grow up to a factor of 2 for HL-LHC conditions due to two times larger initial beam intensity.}
\label{t:flux}
\centering
   \def\arraystretch{1.2}
\begin{tabular}{l c}
\hline
\hline
                                & proton flux on target [p/s]    \\
\hline
estimated for the proposed layout & $\rm 7.5 \times 10^{5}$        \\
\hline
ALICE acquisition system        & $\rm 10^{7}$                   \\
\hline
\hline
\end{tabular}
\end{table}
\section{Conclusions and outlook}
This paper summarises the layout for ALICE \mbox{fixed-target} experiments based on crystal-assisted beam halo splitting. We have demonstrated that it can be operated safely without affecting the availability of the LHC for regular beam-beam collisions. The estimated proton flux on target is roughly one order of magnitude away from  the design goal of $\rm 10^{7}$, with a possible improvement resulting from larger intensities of HL-LHC beams, allowing to exploit of the full capabilities of the ALICE detector acquisition system.

\section{Acknowledgements}
We thank for the support provided by members of the ALICE Fixed-Target Project meetings~\cite{A-FT_meetings} and Physics Beyond Colliders Fixed-Target Working Group~\cite{PBC_WG}, especially 
D.~Kikoła, 
L.~Massacrier and  M.~Ferro-Luzzi. 

This research has received funding from the European Union’s Horizon 2020
research and innovation programme, project number: 101003442. This research was also funded in part by the National Science Centre, Poland, project number: 2021/43/D/ST2/02761.



\begin{thebibliography}{10}
\providecommand{\url}[1]{#1}
\csname url@samestyle\endcsname
\providecommand{\newblock}{\relax}
\providecommand{\bibinfo}[2]{#2}
\providecommand{\BIBentrySTDinterwordspacing}{\spaceskip=0pt\relax}
\providecommand{\BIBentryALTinterwordstretchfactor}{4}
\providecommand{\BIBentryALTinterwordspacing}{\spaceskip=\fontdimen2\font plus
\BIBentryALTinterwordstretchfactor\fontdimen3\font minus
  \fontdimen4\font\relax}
\providecommand{\BIBforeignlanguage}[2]{{%
\expandafter\ifx\csname l@#1\endcsname\relax
\typeout{** WARNING: IEEEtran.bst: No hyphenation pattern has been}%
\typeout{** loaded for the language `#1'. Using the pattern for}%
\typeout{** the default language instead.}%
\else
\language=\csname l@#1\endcsname
\fi
#2}}
\providecommand{\BIBdecl}{\relax}
\BIBdecl

\bibitem{LHC}
\BIBentryALTinterwordspacing
O.~S. Brüning \emph{et~al.}, \emph{{LHC Design Report}}, ser. CERN Yellow
  Reports: Monographs.\hskip 1em plus 0.5em minus 0.4em\relax Geneva: CERN,
  2004.  \url{http://cds.cern.ch/record/782076}
\BIBentrySTDinterwordspacing

\bibitem{A-FT_proposal}
\BIBentryALTinterwordspacing
F.~Galluccio \emph{et~al.}, ``{Physics opportunities for a fixed-target
  programme in the ALICE experiment}'', Apr 2019.
  \url{https://cds.cern.ch/record/2671944}
\BIBentrySTDinterwordspacing

\bibitem{ALICE}
\BIBentryALTinterwordspacing
K.~Aamodt \emph{et~al.}, ``{The ALICE experiment at the CERN LHC. A Large Ion
  Collider Experiment}'', \emph{JINST}, vol.~3, p. S08002. 259 p, 2008, also
  published by CERN Geneva in 2010.  \url{https://cds.cern.ch/record/1129812}
\BIBentrySTDinterwordspacing

\bibitem{PBC-report}
\BIBentryALTinterwordspacing
C.~Barschel \emph{et~al.}, ``{LHC fixed target experiments}'', CERN, Geneva,
  Tech. Rep., Mar 2019.  \url{https://cds.cern.ch/record/2653780}
\BIBentrySTDinterwordspacing

\bibitem{AFTER1}
\BIBentryALTinterwordspacing
C.~Hadjidakis \emph{et~al.}, ``A fixed-target programme at the {LHC}: Physics
  case and projected performances for heavy-ion, hadron, spin and astroparticle
  studies'', \emph{Physics Reports}, vol. 911, pp. 1--83, 2021.
  \url{https://www.sciencedirect.com/science/article/pii/S0370157321000405}
\BIBentrySTDinterwordspacing

\bibitem{Daniele_thesis}
\BIBentryALTinterwordspacing
D.~Mirarchi, ``{Crystal collimation for LHC}'', 2015.
  \url{https://cds.cern.ch/record/2036210}
\BIBentrySTDinterwordspacing

\bibitem{cry_book}
V.~M. Biryukov, Y.~A. Chesnokov, and V.~I. Kotov, \emph{{Crystal channeling and
  its application at high-energy accelerators}}, 1997.

\bibitem{cry_channeling1}
A.~M. {Taratin}, ``{Particle channeling in a bent crystal}'', \emph{Physics of
  Particles and Nuclei}, vol.~29, no.~5, pp. 437--462, Sep. 1998.

\bibitem{scandale2019}
W.~Scandale and A.~Taratin, ``Channeling and volume reflection of high-energy
  charged particles in short bent crystals. crystal assisted collimation of the
  accelerator beam halo'', \emph{Physics Reports}, vol. 815, pp. 1--107, 2019.

\bibitem{scandale2022}
W.~Scandale \emph{et~al.}, ``Feasibility of crystal-assisted collimation in the
  cern accelerator complex'', \emph{International Journal of Modern Physics A},
  vol.~37, no.~13, p. 2230004, 2022.

\bibitem{cry_coll1}
\BIBentryALTinterwordspacing
W.~Scandale \emph{et~al.}, ``{Observation of channeling for 6500 GeV/c protons in the crystal
  assisted collimation setup for LHC}'', \emph{Physics Letters B}, vol. 758,
  pp. 129--133, 2016.
  \url{https://www.sciencedirect.com/science/article/pii/S0370269316301514}
\BIBentrySTDinterwordspacing

\bibitem{cry_coll3}
\BIBentryALTinterwordspacing
{Mirarchi, D.}, {Hall, G.}, {Redaelli, S.}, and {Scandale, W.}, ``{Design and
  implementation of a crystal collimation test stand at the Large Hadron
  Collider}'', \emph{Eur. Phys. J. C}, vol.~77, no.~6, p. 424, 2017.
  \url{https://doi.org/10.1140/epjc/s10052-017-4985-4}
\BIBentrySTDinterwordspacing

\bibitem{cry_coll4}
\BIBentryALTinterwordspacing
{Redaelli, S.} \emph{et~al.}, ``First observation of ion beam channeling in
  bent crystals at multi-{TeV} energies'', \emph{Eur. Phys. J. C}, vol.~81,
  no.~2, p. 142, 2021.  \url{https://doi.org/10.1140/epjc/s10052-021-08927-x}
\BIBentrySTDinterwordspacing

\bibitem{HB2021}
\BIBentryALTinterwordspacing
M.~Patecki, A.~Fomin, D.~Kiko\l{}a, D.~Mirarchi, and S.~Redaelli, ``{Status of
  Layout Studies for Fixed-Target Experiments in Alice Based on
  Crystal-Assisted Halo Splitting}'', \emph{Proceeding to the 64th ICFA
  Advanced Beam Dynamics Workshop, MOP26}.
  \url{https://doi.org/10.18429/JACoW-HB2021-MOP26}
\BIBentrySTDinterwordspacing

\bibitem{PBC}
\BIBentryALTinterwordspacing
Kickoff meeting for the Physics Beyond Collider study, March 2016.
  \url{https://indico.cern.ch/event/523655}
\BIBentrySTDinterwordspacing

\bibitem{PBC_fran}
\BIBentryALTinterwordspacing
F.~Galluccio, ``{UA9} proposal for beam splitting in {IR2}'', {FTE@LHC and
  NLOAccess STRONG 2020 joint kick-off meeting}, 8.11.2019.
  \url{https://indico.cern.ch/event/853688/contributions/3620725/}
\BIBentrySTDinterwordspacing

\bibitem{PBC_alex}
\BIBentryALTinterwordspacing
A.~Fomin, ``{Updates on IP2 FT layouts}'', {{21st meeting of the PBC-FT working
  group}, 16.12.2020}.  \url{https://indico.cern.ch/event/981210/}
\BIBentrySTDinterwordspacing

\bibitem{Daniele_LHCb}
D.~Mirarchi, A.~Fomin, S.~Redaelli, and W.~Scandale, ``{Layouts for
  fixed-target experiments and dipole moment measurements of short-lived
  baryons using bent crystals at the LHC}'', \emph{The European Physical
  Journal C}, vol.~80, 10 2020.

\bibitem{HL-LHC}
\BIBentryALTinterwordspacing
G.~Apollinari \emph{et~al.}, \emph{{High-Luminosity Large Hadron Collider
  (HL-LHC): Technical Design Report V. 0.1}}, ser. CERN Yellow Reports:
  Monographs.\hskip 1em plus 0.5em minus 0.4em\relax Geneva: CERN, 2017.
  \url{https://cds.cern.ch/record/2284929}
\BIBentrySTDinterwordspacing

\bibitem{coll_system}
\BIBentryALTinterwordspacing
R.~W. Assmann \emph{et~al.}, ``{The final collimation system for the LHC}'', p.
  4 p, Jul 2006, revised version submitted on 2006-09-15 14:36:59.
  \url{http://cds.cern.ch/record/972336}
\BIBentrySTDinterwordspacing

\bibitem{HL-LHC_coll}
\BIBentryALTinterwordspacing
S.~Redaelli, R.~Bruce, A.~Lechner, and A.~Mereghetti, ``{Chapter 5: Collimation
  system}'', pp. 87--114, 2020.  \url{https://cds.cern.ch/record/2750434}
\BIBentrySTDinterwordspacing

\bibitem{optics_v1.5}
\BIBentryALTinterwordspacing
{LHC optics repository}.
  \url{https://abpdata.web.cern.ch/abpdata/lhc_optics_web/www/hllhc15/}
\BIBentrySTDinterwordspacing

\bibitem{HL-LHC2}
\BIBentryALTinterwordspacing
O.~Aberle \emph{et~al.}, \emph{{High-Luminosity Large Hadron Collider (HL-LHC):
  Technical design report}}, ser. CERN Yellow Reports: Monographs.\hskip 1em
  plus 0.5em minus 0.4em\relax Geneva: CERN, 2020.
  \url{https://cds.cern.ch/record/2749422}
\BIBentrySTDinterwordspacing

\bibitem{dump1}
\BIBentryALTinterwordspacing
E.~Carlier, ``{Safe Injection into the LHC}'', 2003.
  \url{https://cds.cern.ch/record/642470}
\BIBentrySTDinterwordspacing

\bibitem{dump2}
\BIBentryALTinterwordspacing
R.~Filippini, E.~Carlier, L.~Ducimetière, B.~Goddard, and J.~Uythoven,
  ``{Reliability Analysis of the LHC Beam Dumping System}'', 2005.
  \url{https://cds.cern.ch/record/841092}
\BIBentrySTDinterwordspacing

\bibitem{target_design}
\BIBentryALTinterwordspacing
K.~Pressard, ``{Solid target design for ALICE}'', {Joint workshop
  "GDR-QCD/QCD@short distances and STRONG2020/PARTONS/FTE@LHC/NLOAccess"},
  2.06.2021.
  \url{https://indico.ijclab.in2p3.fr/event/7201/contributions/22534/}
\BIBentrySTDinterwordspacing

\bibitem{madx}
\BIBentryALTinterwordspacing
H.~Grote and F.~Schmidt, ``{MAD-X: An Upgrade from MAD8}'', no.
  CERN-AB-2003-024-ABP, p. 4 p, May 2003.
  \url{http://cds.cern.ch/record/618496}
\BIBentrySTDinterwordspacing

\bibitem{sixtrack1}
\BIBentryALTinterwordspacing
G.~Ripken and F.~Schmidt, ``{A symplectic six-dimensional thin-lens formalism
  for tracking}'', CERN, Geneva, Tech. Rep. CERN-SL-95-12. CERN-SL-95-12-AP.
  DESY-95-063, Apr 1995.  \url{https://cds.cern.ch/record/281283}
\BIBentrySTDinterwordspacing

\bibitem{sixtrack2}
\BIBentryALTinterwordspacing
\emph{{ICFA Mini-Workshop on Tracking for Collimation in Particle Accelerators:
  CERN, Geneva, Switzerland 30 Oct 2015. ICFA Mini-Workshop on Tracking for
  Collimation in Particle Accelerators}}, CERN.\hskip 1em plus 0.5em minus
  0.4em\relax Geneva: CERN, Dec 2018.  \url{https://cds.cern.ch/record/2646800}
\BIBentrySTDinterwordspacing

\bibitem{sixtrack3}
\BIBentryALTinterwordspacing
A.~Mereghetti \emph{et~al.}, ``{Sixtrack-Fluka Active Coupling for the Upgrade
  of the SPS Scrapers}'', \emph{Conf. Proc.}, vol. C130512, p. WEPEA064, 2013.
  \url{https://cds.cern.ch/record/1636191}
\BIBentrySTDinterwordspacing

\bibitem{A-FT_meetings}
\BIBentryALTinterwordspacing
{ALICE Fixed-Target Project meetings}.
  \url{https://indico.cern.ch/category/11595/}
\BIBentrySTDinterwordspacing

\bibitem{PBC_WG}
\BIBentryALTinterwordspacing
{Physics Beyond Colliders Working Group}.
  \url{https://indico.cern.ch/category/8815/}
\BIBentrySTDinterwordspacing

\end{thebibliography}
\end{document}